\documentclass[10pt,conference]{IEEEtran}
\makeatletter
\IEEEoverridecommandlockouts

\def\ps@headings{%
\def\@oddhead{\mbox{}\scriptsize\rightmark \hfil \thepage}%
\def\@evenhead{\scriptsize\thepage \hfil \leftmark\mbox{}}%
\def\@oddfoot{}%
\def\@evenfoot{}}
\makeatother 

\usepackage{color}
\usepackage{graphicx}               % For using graphics like figures
\usepackage{algorithm}
\usepackage{algpseudocode}
\usepackage[cmex10]{amsmath}
\usepackage{soul}
\usepackage{xcolor}

\usepackage{mathtools}
\usepackage[%  
    colorlinks=true,
    pdfborder={0 0 0},
    linkcolor=black
]{hyperref}
\hypersetup{citecolor=black}
\ifCLASSINFOpdf
\else
\fi

\usepackage{algorithm}

\usepackage{algpseudocode}
\usepackage{caption}

\usepackage{url}
\usepackage{tikz}
\usepackage{amsmath}
\usepackage{amsthm}
\usepackage{amssymb}
\usepackage{mathtools, nccmath}
\usepackage{wrapfig}
\usepackage{comment}

\usepackage{blindtext}

\usepackage{graphicx}
\usepackage{xspace}

\usepackage{array}
\usepackage{ragged2e}
\newcolumntype{P}[1]{>{\RaggedRight\hspace{0pt}}p{#1}}
\newcolumntype{X}[1]{>{\RaggedRight\hspace*{0pt}}p{#1}}

\usepackage{tcolorbox}

\usetikzlibrary{arrows,shapes,positioning,shadows,trees,mindmap}
\usepackage[edges]{forest}
\usetikzlibrary{arrows.meta}
\colorlet{linecol}{black!75}
\usepackage{xkcdcolors} % xkcd colors

\usepackage{tikz}
\usetikzlibrary{backgrounds}
\usetikzlibrary{arrows,shapes}
\usetikzlibrary{tikzmark}
\usetikzlibrary{calc}
\usepackage{soul}

\begin{document}

%\title{Efficient Privacy Preserving Aggregation for IoT}
\title{Multi-Party Computation in IoT for Privacy-Preservation}
% % author names and affiliations
% % use a multiple column layout for up to three different

% \author{\IEEEauthorblockN{Himanshu Goyal}
% \IEEEauthorblockA{Indian Institute of Technology Bhubaneswar\\
% hg11@iitbbs.ac.in} \and \IEEEauthorblockN{Sudipta Saha}\IEEEauthorblockA{Indian Institute of Technology Bhubaneswar\\ sudipta@iitbbs.ac.in} 
%}
% \author{Vitaly Surazhsky\thanks{Department of Computer Science,
%         Technion---Israel Institute of Technology,
%         Technion City, Haifa 32000, \underline{Israel}}        
%     \and    
%     Yossi Gil\thanks{Department of Computer Science,
%         Technion---Israel Institute of Technology,
%         Technion City, Haifa 32000, \underline{Israel}}}%

\author{\IEEEauthorblockN{Himanshu Goyal,
Sudipta Saha}
\IEEEauthorblockA{Indian Institute of Technology, Bhubaneswar. Email: \{\textit{hg11}, \textit{sudipta}\}@iitbbs.ac.in}}


% \author{
% \IEEEauthorblockN{Jagnyashini Debadarshini}
% \IEEEauthorblockA{IIT Bhubaneswar\\
% jd12@iitbbs.ac.in} 
% \and 
% \IEEEauthorblockN{Sudipta Saha}
% \IEEEauthorblockA{IIT Bhubaneswar\\
% sudipta@iitbbs.ac.in} 

% }
% \IEEEoverridecommandlockouts
% \IEEEpubid{\makebox[\columnwidth]{978-1-5386-5541-2/18/\$31.00~\copyright2018 IEEE \hfill} \hspace{\columnsep}\makebox[\columnwidth]{ }}
\maketitle
%\IEEEpubidadjcol


% \fi
\IEEEpeerreviewmaketitle
\vspace{-0.2cm}
\begin{abstract}
Preservation of privacy has been a serious concern with the increasing use of IoT-assisted smart systems and their ubiquitous smart sensors. To solve the issue, the smart systems are being trained to depend more on aggregated data instead of directly using raw data. However, most of the existing strategies for privacy-preserving data aggregation, either depend on computation-intensive \textit{Homomorphic Encryption} based operations or communication-intensive collaborative mechanisms. Unfortunately, none of the approaches are directly suitable for a resource-constrained IoT system. In this work, we leverage the \textit{concurrent-transmission}-based communication technology to efficiently realize a Multi-Party Computation (MPC) based strategy, the well-known \textit{Shamir's Secret Sharing} (SSS), and optimize the same to make it suitable for real-world IoT systems. 

\end{abstract}

%\vspace{-0.1cm}
% \begin{IEEEkeywords}
% \normalfont{\textit{Data Privacy, Distributed computing, Secure Multi-Party Computation, Synchronous transmission, Capture effect, Constructive Interference.}}
% \end{IEEEkeywords}

\vspace{-0.2cm}
\section{Introduction}
\vspace{-0.05cm}

With the increasing use of smart-systems ubiquitous availability of sensitive data has become a matter of serious concern. \textit{Privacy-Preserving Data-Aggregation} (PPDA) has gained a lot of research attention in recent days \cite{survey1,survey2} as a potential solution to this issue. However, most of the existing PPDA solutions rely on highly computation-intensive \textit{Homomorphic Encryption} (HE). Hence, they mostly do not fit with resource-constrained IoT systems. \textit{Secure-Multi Party Computation} (SMPC) based strategies, in contrast, approach a collaborative solution for PPDA and hence, depend comparatively less on computation, while heavily using communication/data sharing among the entities. However, the communication hardware being the most energy-hungry unit, the IoT devices always try minimization of their communication requirements too in order to have sustained life. Thus, none of the existing solutions for PPDA is directly applicable to resource-constrained IoT systems.

Efficient communication in low-power IoT systems has been an active research area. In recent works, \textit{Concurrent-Transmission} (CT) based communication in IoT/WSN \cite{sync_survey} has got quite good popularity because of its ability to achieve high reliability and simultaneously consume very less time. Different standard communication patterns including one-to-many, many-to-one as well as many-to-many have been quite successfully realized using CT. In the current work, we approach taking the help of CT to efficiently realize the communication-intensive component of SMPC. In particular, in this work, we make an endeavor to efficiently realize the well-known \textit{Shamir Secret Sharing} (SSS) \cite{sss} strategy with the help of a CT-based data sharing protocol MiniCast \cite{minicast}. 

\section{Background \& Design}
\label{sec:background}
\vspace{-0.045cm}
In the following, we first briefly explain SSS and MiniCast, and subsequently, we explain the main design considerations to optimally integrate the two with each other.

\textbf{Shamir-Secret Sharing (SSS)}: SSS achieves PPDA using polynomial interpolation over finite fields in a semi-honest adversarial setting. Every node $n_i$, assumes a $k$-degree polynomial $P_{i}$ with coefficients $c_{1,i}$, $c_{2,i}$,\ldots, $c_{k-1,i}$, where $c_{k-1,i} = S_i (= P_{i}(0))$, i.e., the secret value of node $n_i$. The aggregation process happens in two steps. During aggregation, every node evaluates its own polynomial at a set of $n$ number of public-points and subsequently shares these values to specific nodes through secure channels. This is referred to as the \textit{sharing phase}. Every node is designated for a specific public-point based on the ID of the node. Each node then locally sums up the share values received from different nodes. Finally, the nodes re-share these sum values with each other. This is referred to as \textit{reconstruction-phase}. At this stage, a node can use any set of $k+1$ values to reconstruct a final polynomial $(P_{s}(x)$ (using \textit{Lagrange Interpolation} technique) which is the sum of the polynomials hold by all the nodes, i.e., $P_{s}(x) = \sum_{i=1}^{n} P_{i}(x)$. The aggregation of the secret value is obtained as the constant term of $P_{s}(x)$, i.e., $P_{s}(0) = \sum_{i=1}^{n} S_{i}$.

\textbf{MiniCast}: In SSS, a very frequent and important job is the dissemination of data from many source nodes to all the nodes which we carry out using the CT-based protocol MiniCast\cite{minicast}. MiniCast extends the functionality of the CT-based pioneering protocol Glossy\cite{glossy}. Fundamentally, it enables multiple instances of the Glossy-based flood to run together in an interspersed manner to realize all-to-all/many-to-many data sharing. To accommodate multiple floods within the same time frame, the protocol arranges all the transmissions in the chain of packets, i.e., as a sequence of packets, based on a TDMA schedule. The process starts from a designated initiator node. Subsequently, the first-hop neighbors of the initiator transmit their packets which in turn trigger the transmission from the second hop. The parameter NTX in MiniCast defines the number of times a node transmits a full chain of packets. 

\textbf{MiniCast hosting SSS}: The two rounds of SSS directly maps to two rounds of MiniCast. In sharing phase, in order to enable a node $n_i$ to share $n$ number of evaluated values destined for distinct nodes, the chain size is extended to contain $n^2$ sub-slots (packets) where the packet transmitted in each sub-slot is encrypted by a key which is assumed to be already shared with the destination node during the bootstrapping phase. In the reconstruction phase, the nodes share the sum values for different public points, hence, the chain size of $n$ is enough. However, chain size in the sharing phase is $O(n^2)$. 

\section{Scalable Shamir Secret Sharing}

The degree of the polynomial being used in SSS decides the collusion threshold and overall protection of privacy. In the best case, the degree can be set as the number of nodes. However, for many practical purposes setting a lower degree is also enough. Therefore, we observe that the use of low-degree polynomial can bring multi-fold optimization. First, for a low-degree polynomial, the overall chain size in the sharing phase will decrease substantially. Second, to support a low-degree polynomial the sharing phase may run for a duration that would be enough to reach out only to the necessary number of neighbors, instead of attaining full network coverage. Moreover, having a low-degree polynomial also introduces an advantage of a higher degree of fault tolerance. In particular, when a degree $k$ polynomial is used where $k<n$, in the reconstruction phase even the final polynomial can be formed by combining any $k+1$ sum values. This alleviates the need for strict all-to-all sharing of data in the reconstruction phase also making the protocol fault-tolerant. 

The parameter NTX in MiniCast mainly controls the degree of coverage/reliability the dissemination process achieves. For sufficiently large NTX, a node gets the data from all the other nodes in the network. However, MiniCast shows a very stable and consistent behavior even for low NTX, although fails to receive the full network coverage. A node successfully receives the data from its neighbor within a certain perimeter depending on the exact value of NTX. The behavior is quite non-linear, i.e., with a short increase in NTX, a large amount of data becomes available in a node, while it takes a comparatively higher time (NTX) to have the full network coverage. 

We exploit the above-mentioned behavior of MiniCast to optimize the sharing phase of SSS. In summary, we assume that the degree of the polynomial ($p$) is low. To improve the performance, not only the chain size in sharing phase is trimmed but also, the data sharing process is executed for quite a low NTX value. However to achieve this, in the bootstrapping phase every node is assumed to take note of which neighbor is reachable at what NTX value. Using this information, the chain in the sharing phase is constructed in a way where a node shares evaluation values and encrypts them for a few known pre-determined neighbors. The process completes fast with low NTX and enters the reconstruction phase to complete the process.

\vspace{-0.2cm}
\section{Evaluation}
\label{sec:results}

The proposed strategy is implemented in Contiki OS for nRF52840. The naive implementation of SSS is referred to as S3 while the scalable version is referred to as S4. We execute S3 and S4 in two publicly available testbeds FlockLab \cite{flocklab} and DCube\cite{dcube} having 26 and 45 nRF devices. In the sharing phase, each packet is encrypted using AES-128 while the reconstruction phase runs in plane text. We experiment with the different number of source nodes. We assume a polynomial of degree nearly equal to \(\ \lfloor \frac{n}{3} \rfloor\), $n$ being the total number of nodes in both S3 and S4. Each experiment is repeated for 2000 iterations. The value of NTX as 6 and 5 are found to be enough for sharing the data within the necessary number of neighbors in the sharing phase in FlockLab and DCube respectively.

\begin{figure}
\vspace{-0.2cm}
    \centering
    \includegraphics[scale=0.8]{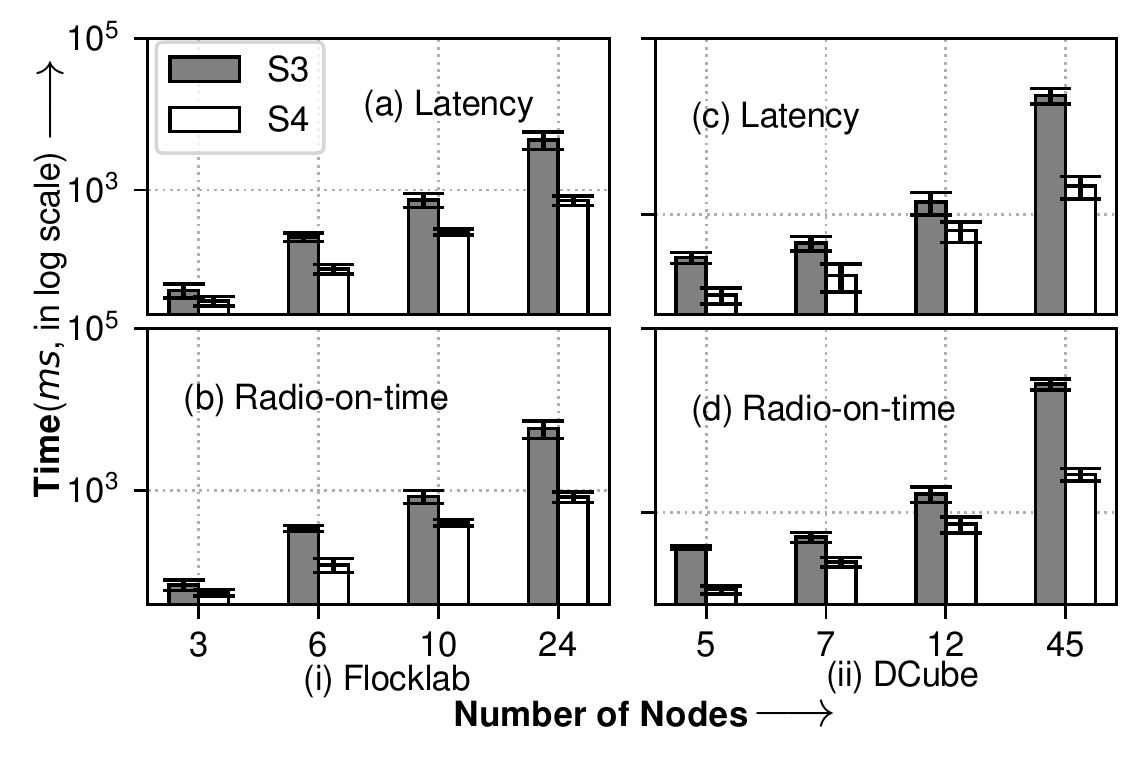}
    \vspace{-0.9cm}
    \caption{Performance comparison between S3 and S4 in FlockLab (26 nodes) and DCube (45 nodes).}
    \label{fig:result}
    \vspace{-0.7cm}
\end{figure}

Two standard metrics \textbf{Latency} (time required to obtain the final aggregation in each node) and \textbf{Radio-on time} (time necessary to complete the communication process in a round) are used for comparing the performance of S3 and S4 as depicted in Fig. \ref{fig:result}. For complete network, it can be seen that S4 achieves private aggregation at least 6\(\times\) faster and consuming 7\(\times\) lesser radio-on time in FlockLab and 9\(\times\) faster and consuming 10\(\times\) lesser radio-on time in DCube compared to S3. Note that further improvement in the latency and radio-on time would be visible in S4 compared to S3 for an even lesser degree of the polynomial used.

\section{Conclusion}
\label{sec:conclusion}

In this work, we show a possible lightweight and efficient realization of the well-known Multi-Party Computation-based strategy SSS for resource-constrained IoT systems. To make it time and energy-efficient, we introduced a couple of optimizations over the naive strategy based on the observation that for many practical cases we do not need the highest degree of privacy protection or collision resistance.

\vspace{-0.2cm}
\bibliographystyle{IEEEtran}
\bibliography{references}

\end{document}